\documentclass[10pt,journal]{IEEEtran}

\usepackage{amsmath,epsfig}

\ifCLASSINFOpdf

\else

\fi

\begin{document}
\onecolumn

\title{Interactive multiview video system with  non-complex  navigation at the decoder}

\author{\IEEEauthorblockN{Thomas~Maugey 
        and~Pascal~Frossard}\\
\IEEEauthorblockA{Signal Processing Laboratory (LTS4)\\
\' Ecole Polytechnique F\'ed\'erale de Lausanne (EPFL), Lausanne, Switzerland\\
\{thomas.maugey,pascal.frossard\}@epfl.ch}
}

\maketitle

\begin{abstract}
Multiview video with interactive and smooth view switching at the receiver is a challenging application with several issues in terms of effective use of storage and bandwidth resources, reactivity of the system, quality of the viewing experience and system complexity.  The classical decoding system for generating virtual views first projects a reference or encoded frame to a given viewpoint and then fills in the holes due to potential occlusions. This last step still constitutes a  complex operation with specific software or hardware at the receiver and  requires a certain quantity of information from the neighboring frames for insuring consistency between the virtual images.  In this work we propose a new approach that shifts most of the burden due to interactivity from the decoder to the encoder, by anticipating the navigation of the decoder and sending auxiliary information that guarantees  temporal and interview consistency. This leads to an additional cost in terms of transmission rate and storage, which we minimize by using optimization techniques based on the user behavior modeling. We show by experiments that the proposed system represents a valid solution for interactive multiview systems with classical decoders.

\end{abstract}


\IEEEpeerreviewmaketitle

\section{Introduction}
Providing a three dimensional impression in multimedia applications is a challenging task that requires to properly study the sender/receiver interactions. The end-to-end system (\emph{i.e.,} with capture, description, coding, transmission, decoding, display, see for example \cite{Benzie_P_2007_tcsvt_sur_tdtvtt,Holliman_N_2011_tb_thr_ddraa}) sensibly varies depending on the target applications. 
The coding of multiview sequences have been widely explored in the scenario where the whole set of frames for all views is transmitted together to servers, edge-servers or client directly. In this configuration, increasing the coding efficiency leads to better exploitation of the interdependencies between the frames. This could be done by extending the motion estimation to inter-view prediction  \cite{Merkle_P_2007_tcsvt_eff_psmvc,Vetro_A_2011_pieee_ove_smvcehms}. In some approaches, the inter-frame correlation is exploited using the geometry of the scene, \emph{e.g.,} with depth images \cite{Muller_K_2011_pieee_tdv_rudm}. Therefore, novel algorithms have been proposed lately to improve the depth information compression \cite{Oh_H_2006_hts_bdmscumictv,Morvan_Y_2007_picip_dep_icbrdoqdtmi} and to smartly balance the  rate dedicated to texture and geometry information \cite{Merkle_P_2007_picip_mvvpdrc,Merkle_P_2008_tdtvc_eff_dcmrq}.

The delivery of all the frames is however unadapted to interactive systems. Indeed, in this situation, the user only needs to receive the requested views and not the whole set of frames. Therefore, one needs to define alternatives to the classical prediction structure of MVC, which is efficient only if all the frames are transmitted, but very limited if only a selection of the images are sent to a receiver. This can be achieved by limiting the dependencies in the multiview video coding algorithm, or by sending additional information to help navigation at the decoder. There exists a compromise between the level of interactivity (system delay, image quality, frame rate, number of available views, etc.) and the cost of this interactivity service (bandwidth or storage size). As this application typically targets simple devices such as mobile, TV decoder, personal computer, they should not involve too much complexity at the decoder side. However the majority of the existing interactive systems do not pay attention to the computational cost of their decoding algorithm for example in the synthesis of virtual views. In contrary to what is stated in the majority of the papers related to interactive multiview systems, the design of synthesis algorithms is not obvious. The frame reconstruction techniques need information  from the neighboring frames for guaranteeing consistency between the frames; this is not handled by the existing systems although  generating good quality synthesized views and smooth transitions between the cameras creates a "look around effect" necessary to give the impression of immersion in the scene when the stereoscopic display is not available at the receiver, which is our assumption in this work.

In this paper, we propose a new system for multiview video transmission, which enables both a low complex interactivity and acceptable temporal and inter-view consistency.  This original scheme is based on the idea of transmitting additive information in order to help the decoding process at the receiver. In the classical video coding schemes, the additive information (or residual information) is usually transmitted to enhance the decoding quality. Here we propose to study the cost and the efficiency of this residual information to decrease the complexity of interactive decoders. Our focus is to study the balance between rates, navigation capabilities and complexity in interactive multiview systems. For that purpose,  we build a complete scheme that provides a very satisfying interactivity with low complexity and good viewing experience. We propose to construct and code residual frame information at the server which is used for interactive navigation at the decoder. We define a rate-distortion effective encoding of this information using the user behavior models. We finally show by extensive experiments that our scheme is a valid solution for low complexity interactive navigation systems, and presents an effective trade-off between interactivity and system resources. 

The paper is organized as follows. We first introduce in Sec.~\ref{sec:lowDecCompl} the original idea of our system that consists in encoding some additive residuals (called e frames) in order to help the decoder to reduce the calculation costs due to navigation. In Sec.~\ref{sec:propSyst}, we detail the complete system that permits the transmission of the multiview video and the e frames. Then, we propose in Sec.~\ref{sec:RA} some rate-distortion optimization of our system. Finally we show in Sec.~\ref{sec:exp} the performance of our system with extensive experiments.

\section{Low complexity view synthesis}\label{sec:lowDecCompl}
\subsection{Framework}
The target of the proposed system is to deliver to a receiver (or to multiple receivers) a video sequence acquired in a multiview system with a fixed number of color+depth cameras. In addition the user should be able to choose the view and to change the viewpoint. In other words, the receiver only displays a 2D image on a classical video decoder. This image corresponds to one viewing angle in the multiview framework, and the user has the possibility to ask the server to change the viewpoint. The system thus has the objectives of minimizing the delay between the request and the actual viewpoint modification, of providing a high visual quality, of enabling the user to choose between a large number of viewpoints, of minimizing the required rate, and of finally keeping the decoding complexity at a reasonable level. This last requirement leads to non-classical system design, where the server has to prepare additional information used for interactivity.

\subsection{Requirements posed by interactivity}
For good visual quality, an interactive multiview system has to enable smooth transitions between the different views requested by the user, which is motivated by the need of immersion in the scene. This is called the \emph{look around effect} \cite{Muller_K_2011_pieee_tdv_rudm} and requires a very high number of available views at the decoder. On the one hand it is important to propose a large amount of neighbor views to the user in order to satisfy  this desire of immersion; on the other hand increasing the number of cameras is quite costly in terms of hardware.  Smooth navigation thus comes through the generation of virtual views at the decoder.
Usually, a virtual view synthesis (VVS) algorithm is composed of two steps: i) prediction and ii) error concealment. The prediction step consists in estimating the displacement of each pixel from the reference image to the target virtual view, using depth information. This operation is well described in \cite{Muller_K_2011_pieee_tdv_rudm, Muller_K_2008_jivp_vie_satdvs,Tian_D_2009_pspie_vie_sttdv} or \cite{website_VVS}. The general idea of the process is to first project a pixel in the image plane coordinate (2D), then to the camera coordinate (3D) using depth information and intrinsic camera parameters, and finally to the world coordinate (3D) using the extrinsic camera parameters. In a second part, the inverse process is performed, and the pixel is projected from the world coordinate to its position in the virtual view with the target camera parameters. If two reference cameras are used for VVS, the above projection is performed once for every camera; a fusion algorithm merges both projection results by considering distances to the reference cameras. 
This process leaves some holes in the image due to occlusions.  They are usually filled by applying an inpainting algorithm. Inpainting algorithms \cite{Criminisi_A_2004_tip_reg_forebii} have been generally used in order to conceal image areas affected by manual object removal or any other type of local degradation. Some works have proposed  adaptation of  inpainting techniques to the occlusion filling problem \cite{Daribo_I_2010_mmsp_dep_aiinvs,Oh_KJ_2009_ppcs_hol_fmudbivsfvt3dv}; they use depth and neighboring view information in order to generate estimations that lead to time and view consistency in the reconstructed images.

This classical VVS algorithm structure however has two major limitations that are generally not taken into account in the literature. First, the dense projection and the inpainting algorithms are both very complex for a light decoder. Secondly, if the hole filling algorithm does not use any information taken from the neighboring frames, it reconstructs the images without really guaranteeing temporal or inter-view consistency. Yet, it is commonly admitted that flickering effects (due to inconsistency between frames) are very damageable for the visual quality. Instead of relying purely on VVS with received frames, the decoder thus requires some additional information transmitted by the encoder in order to enable a high quality reconstruction with possibly lower computational requirements. Finally, the implementation of an effective interactive system leads to a trade-off between transmission rate, visual quality and computational complexity at the decoder.

\subsection{E frames}

\begin{figure}[!t]

\begin{minipage}{1\linewidth}
\centering
\includegraphics[width=0.8\linewidth]{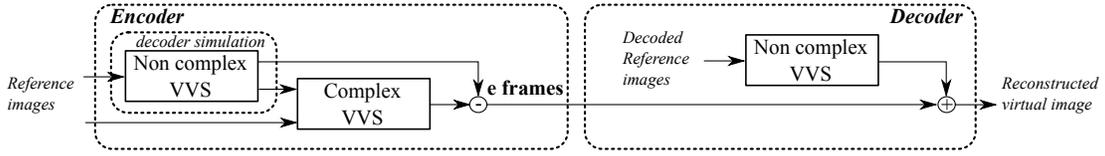}
\center{\footnotesize (a) The e frames are built by estimating the difference between a non complex virtual view synthesis (VVS) and a good quality virtual view.}
\end{minipage}
\begin{minipage}{1\linewidth}
\vspace{0.5cm}
\centering
\includegraphics[width=1\linewidth]{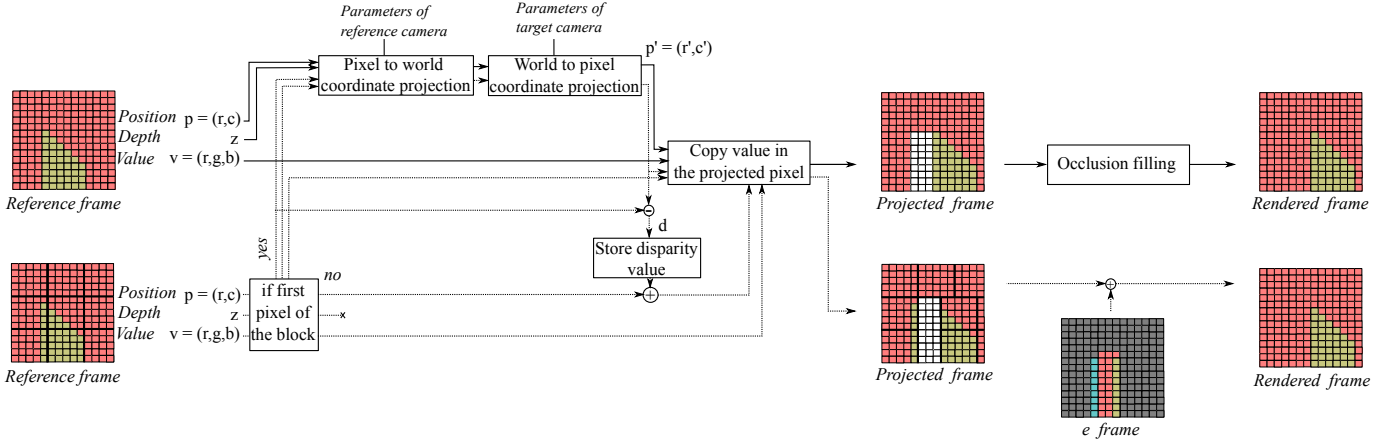}
\center{\footnotesize (b) Difference between the complex (plain arrows) and the non-complex (dashed arrows) Virtual view synthesis (VVS) algorithms performed at the decoder. In the second case the e frames are used to enhance the virtual views.\\}
\end{minipage}

\caption{Description on the e frame generation and their use at the decoder side.}
\label{fig:eFrameGeneration}
\end{figure}

 Based on the observations from the previous section, we propose  to build and transmit auxiliary information in order to help the decoder for the creation of virtual views. This additional information needs to be simple to decode, unlike the hash information streams considered in some other schemes  \cite{Cheung_G_2011_tip_int_ssmvurfs}.  With this additional information, part of the calculation that is usually performed at the user side is shifted to the encoder. We call the additional information as \emph{e frames}, which are built on residual information (see Fig.~\ref{fig:eFrameGeneration}). The idea of transmitting residual information to help the decoder has already been  explored in the literature, but with the purpose of enhancing the decoding efficiency. We can cite for example the classical motion compensation residual in most of the common video codecs \cite{Wiegand_T_2003_tcsvt_ove_hvcs,jmvm}. We also refer to the layered depth video format \cite{Bartczak_B_2011_tb_dis_itdtvpduldvf,Daribo_I_2011_tb_nov_ibldvtdtv}, where correction information resulting from DIBR is also considered. In all these methods the residual information is sent for quality enhancement and not necessarily for lowering the computational requirements at decoder. The residual construction is however similar so that our scheme is  compatible with the classical decoders: the decoder is simulated at the encoder side and the residual information is the difference between the low complexity decoded version without auxiliary information and a ``good quality" version of the signal (Fig.~\ref{fig:eFrameGeneration}~(a)).

\begin{figure}[tph]
\centering
\begin{minipage}[b]{0.48\linewidth}
  \centering
 \centerline{\epsfig{figure=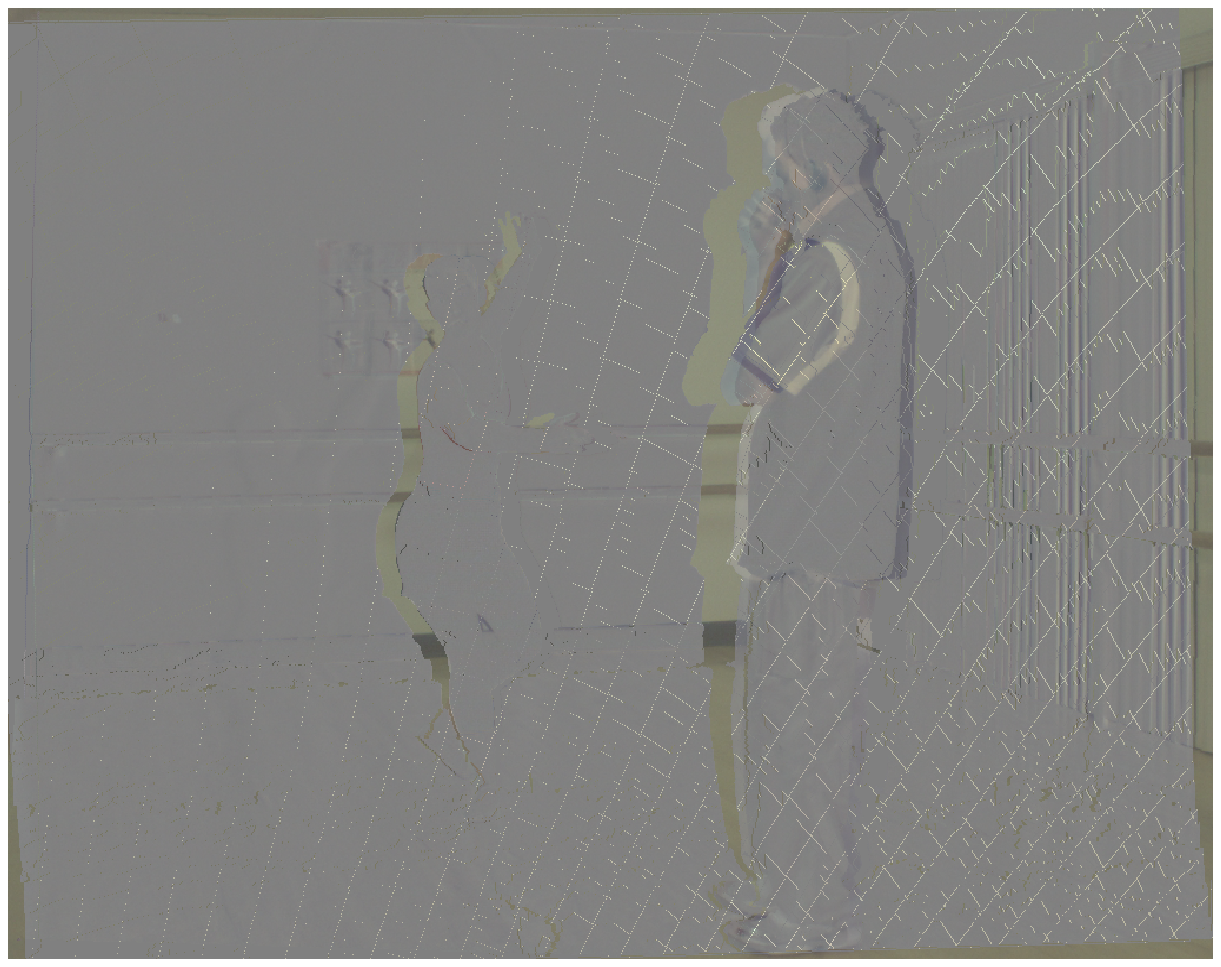,width=0.82\linewidth}}
\centerline{(a)}
\end{minipage}
\begin{minipage}[b]{0.48\linewidth}
  \centering
 \centerline{\epsfig{figure=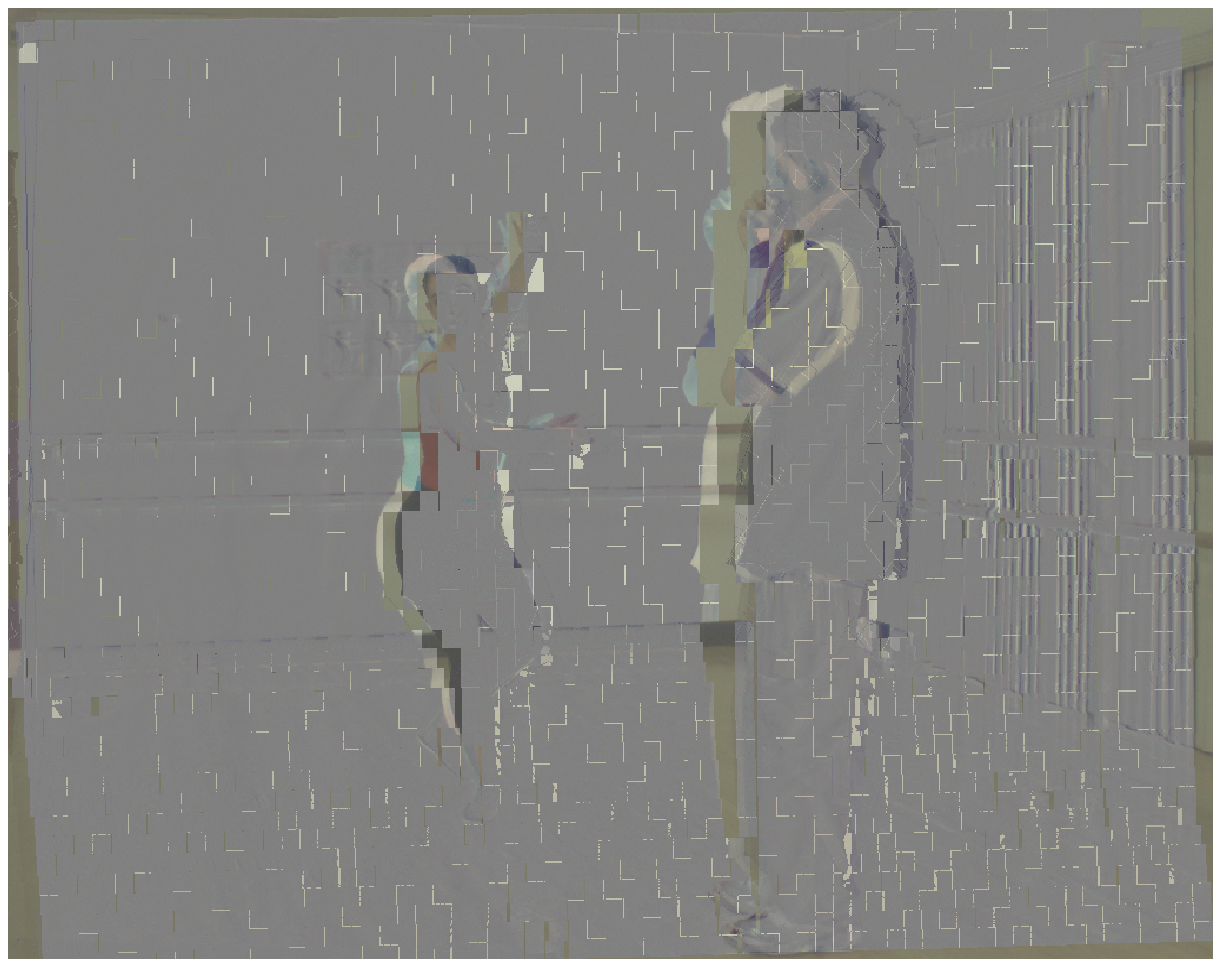,width=0.82\linewidth}}
\centerline{(b)}
\end{minipage}
\caption{Example of transmitted e frame involving (a) the occluded regions (b) the occluded regions and the blocking errors.}
	\label{fig:eframes_example}
\end{figure}

The first idea for complexity reduction at the receiver side is to remove the very complex occlusion filling step from the decoding operation. This is partially done in our previous work \cite{Maugey_T_2011_picip_int_mvsldc}, where the e frames contain the missing parts of the decoded images, as shown in Fig.~\ref{fig:eframes_example}~(a). 
Whereas shifting the occlusion filling operations from the decoder to the encoder has already a significant impact on the decoding complexity, the  projection operation in the construction of virtual views is still too complex for a light hardware:  it involves a pixel-based image compensation that involves several matrix multiplications for each displacement calculation. In the scheme presented in this paper, we thus propose to also reduce the complexity of the projection operation at the decoder. The approach is simple and consists in replacing the pixel precision by a block precision in the projection, where the block size is denoted by $B$\footnote{In this paper, we consider block sizes of $4\times4$, $8\times8$ and $16\times16$.}. In other words, instead of calculating  displacements pixel by pixel with several matrix multiplications, the proposed low complexity decoder performs projection for each block of pixels and uses the same disparity value for every pixel in the block. The dimension and thus the coding rate of the depth maps thus decrease in this case. As the quality of the projection is reduced in block-based approaches, we include in the e frames the resulting estimation error, so that the decoder can reconstruct views of good quality. The e frames thus  contains  the error due to the block-based compensation, as shown in Fig.~\ref{fig:eframes_example}~(b). The overall construction of the e frames is illustrated in details in Fig.~\ref{fig:eFrameGeneration}~(b).

\section{Interactive multiview system}\label{sec:propSyst}
Equipped on the original e-frame idea proposed in the previous section, we present here the general system that offers a non-complex interactivity to the user. 

\begin{table}
\centering
\begin{scriptsize}
\begin{tabular}{c | c | c }
Name & Notation  &  Definition       \\ \hline
GOP size &  $GOP$ & \begin{minipage}{0.7\linewidth} Size of the GOP used to compressed the reference sequences (color and depth) with JSVM \cite{jsvm}  \end{minipage}      \\ \hline
request interval  & $N_T$ &   \begin{minipage}{0.7\linewidth} Interval (in number of frames) between two requests from the user to the server \end{minipage}      \\ \hline
request delay  & $N_D$ & \begin{minipage}{0.7\linewidth} Time (in number of frames) between the request and the effective reception of the demanded frames  \end{minipage}      \\ \hline
Block size & $B$  & \begin{minipage}{0.7\linewidth} size of the blocks used at the projection step of the virtual view synthesis algorithms \end{minipage}      \\ \hline
No switching probability & $p_1$  & \begin{minipage}{0.7\linewidth} probability that the user does not start any right or left switching \end{minipage}      \\ \hline
Continue switching probability & $p_2$  & \begin{minipage}{0.7\linewidth} probability that the user continues his (right or left) switching \end{minipage}      \\ \hline
Stop switching probability & $p_3$  & \begin{minipage}{0.7\linewidth} probability that the user stops his (right or left) switching \end{minipage}      \\ \hline

    \end{tabular}
 \end{scriptsize}
  \caption{System parameters}
  \label{tab:variables}
   \end{table}

\subsection{User interactivity}

For multiview video transmission systems, the purpose of enabling the user to change the viewpoint is twofold. First, it lets the user choose the camera position and angle used to observe a scene. This is especially interesting when watching scenes that contain some localized points of interest such as sport, concert or game events. In that purpose, any kind of interactivity may be considered. In other words, random access or smooth navigation in the multiview content can both be envisaged. On the other hand, interactivity can also provide a sensation of immersion in the scene that could replace complex 3D displays. One classical way of rendering three dimensions to the user is to transmit stereo sequences. The problem is that it requires complex and expensive hardwares (glasses, specific screens, etc). However the 3D impression is also provided by the look around effect due to smooth transitions between the different views \cite{Smolic_A_2005_pieee_int_tdvrct}. This does not require specific hardware on the client's side. It is exactly the objective of our interactive multiview system, where we consider that users might decide to gradually switch views in any direction. For that purpose we also consider the synthetic viewpoints, obtained thanks to the e frames, in order to offer smooth transitions between the captures sequences.

\subsection{Proposed system}\label{sec:propSystem}

\begin{figure}[!t]
\centering
\includegraphics[width=0.4\linewidth]{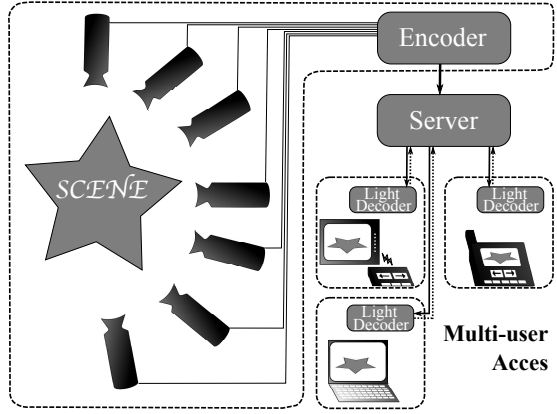}
\caption{ General system structure}
\label{fig:genSystem}
\end{figure}

The general structure of the system is composed by different fonctions: capture, encoding, storage on a server, transmission to the user and decoding, as shown in Fig~\ref{fig:genSystem}. After capture, the datas (color and depth sequences) are compressed and transmitted to a central server called the \emph{main server (MS)}. The server then processes these sequences before storage. Their stored version is a compressed scalable bitstream that the user could access at the quality (or rate) he wants. For this operation, we use the reference scalable video coder described in \cite{jsvm}.  In addition the server generates, codes and stores e frames that correspond to additive information that can be sent to the decoder in order to enhance the virtual view synthesis operation. The e frames described in the previous section reduce the computational power requirements at decoder and increase the quality of the synthesis of virtual views.

At the user side, we assume that a standard video decoder accesses  the information stored on the MS via a networks with feedback channel. On one hand the communication user$\rightarrow$MS enables the server to get some informations about the user navigation, and on the other hand, the communication MS$\rightarrow$user is used to transmit a bitstream that enables the user to navigate between the views. This bitstream corresponds to a group of images called as \emph{set of frame (SoF)}. The communication MS$\leftrightarrow$user depends on two parameters that define the level of interactivity in the system. First, we assume that the interval between two messages between client and server is equivalent to $N_T$ frames,  called as \emph{request interval}; its value is set by the network and can be either fixed or adaptive. This also fixes the interval between two communication from server to client. Second we denote the time spent to transmit the bitstream as \emph{request delay}, $N_D$ expressed in number of frames. Note that a real time interactivity is possible as soon as $N_D~<~N_T$.

\begin{figure}[!t]
\centering
\includegraphics[width=0.6\linewidth]{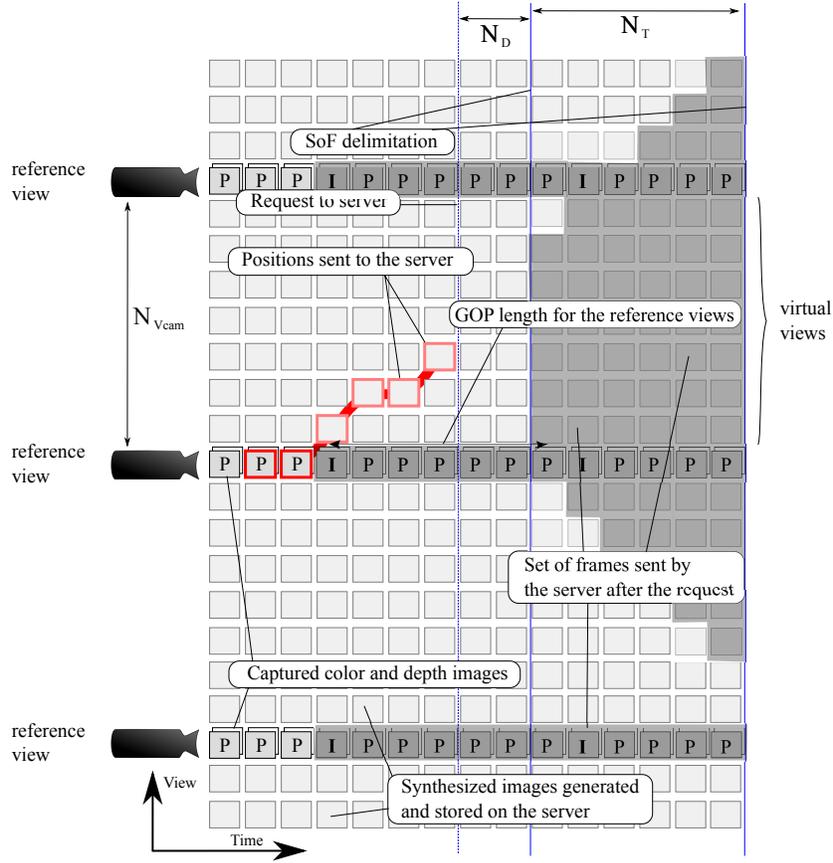}
\caption{ An example of interaction between server and client. The past navigation path is given by the red images.}
\label{fig:propSystem}
\end{figure}

The proposed system allows multiple users with different capabilities to access to multiview content. Indeed the data description on the MS is not specific to one user due to its scalability. The MS only needs to prepare and transmit  data specific to each user as soon as it receives clients' requests. Fig.~\ref{fig:propSystem} shows the detail of the server-client communication process. The highlighted frames correspond to the ones sent to the client after its request. The request happens $N_D$ frames before the effective beginning of the SoF. The SoF contains all the achievable e frames and all the reference frames that are also achievable and/or involved in the e frame generation. Note that, when $N_D$ becomes larger, the number of transmitted e frames obviously increases; this is actually imposed by the network.

\subsection{Server}

We provide now more details about the multiview content that is present at the server. The reference sequences (color and depth) are stored in a H.264 scalable format \cite{jsvm}. The beginning of each GOP (\emph{i.e.}, the first intra frame) is synchronized between the views, in other words the GOP length is fixed and the I frames occurs at the same time in every view. Then the server also stores additional information for low complexity view synthesis, in the form of e frames. The e frame generation process is summarized in Fig~\ref{fig:eFrameGeneration} and is based on two virtual view synthesis (VVS) algorithms. The so-called \emph{non complex VVS} corresponds to the algorithm that is used at the decoder. It is designed such that it involves a low computational power for view synthesis. The \emph{complex VVS} that is implemented at the server uses the output of the non-complex VVS, and the original input images in order to generate a higher quality synthesis of virtual images for navigation. This is considered as the target quality that users should experience. The e frame residual compressed on the server corresponds to the difference between the outputs of the non complex and the complex VVS block. They are used by the clients to replicate the output of the complex VVS algorithm, but with a low complexity decoder.

Note that the VVS algorithms require color and depth information extracted from one or several reference cameras. The number of reference views (\emph{e.g.} one or two) used for e frame generation impacts on the amount of data needed to be transmitted and on the storage size on the server. Using one reference view has the advantage of reducing the need of reference information at the decoder. On the other hand, using two views reduces the size of occlusions and then the rate of the e frames. Moreover, it makes the synthesis problem symmetric and then reduces the number of necessary e frame descriptions. With one reference view, the server has to store two versions per e frames (one per neighboring reference camera), while only one description per e frame is necessary with two reference views. These properties are summarized in Tab.~\ref{tab:RefView}. We consider in this work that two reference frames are used for the view synthesis.

\begin{table}
\centering
\begin{scriptsize}
\begin{tabular}{c | c | c }
 \begin{minipage}{0.6\linewidth} Number of reference view used for e frame generation \end{minipage}  &  One  & Two  \\ \hline
amount of reference transmitted data & \textbf{low} &high \\
e frame size & higher  &  \textbf{lower} \\
decoding complexity & \textbf{lower} & higher \\
number of e frame stored version & two (left + right)  &  \textbf{one} \\
 \hline   \end{tabular}
 \end{scriptsize}
  \caption{Comparison of using one or two reference views for the VVS.}
  \label{tab:RefView}
   \end{table}

\section{Rate-distortion optimized e-frame coding}
\label{sec:RA}

As the server prepares and transmits auxiliary information for offering smooth navigation at the decoder, the storage or bandwidth resource requirements might become important. We propose in this section a method for coding the e frames in a rate-distortion effective way, where we exploit user behavior models.

Let us denote by  $F_{v,t}$  the $t^{th}$ frame of the view $v$ which can be a reference or virtual view. For the following we introduce the notion of \emph{frame popularity}, $P(F_{v,t})$ that corresponds to the probability that the user chooses the view $v$ at time $t$. Under this definition, we have $\sum_{k} P(F_{k,t}) = 1$ as we assume that users look at one frame and only one frame at each instant $t$. In this paper, we further assume, that every frame are \emph{a priori} equiprobable. In other words, we assume that the user may watch the scene from every viewpoint with the same probability.
 Note that this assumption might not be exactly verified in practice because the views can have a different interest depending on the scene content. However, this choice does not limit the generality of our approach and the probability model can be modified without affecting the rest of the system.

For a given user, the frame popularity is however obviously conditioned by the current user position in the multiview context. Indeed, knowing that the user is watching the frame $F_{v',t'}$ obviously impacts on the probability of looking at every $F_{v,t}$ with $t>t'$. To the best of authors' knowledge, it does however not exist any work in the literature that proposes and validates a user navigation model that could help calculating these conditioned probabilities. Thereby, in this work, we propose a simple empiric model that relies on basic observations of user behavior. In other words, we  assume that a good user behavior model is known at the server, but the actual instance of such a model is not critical in our optimization methodology.

\begin{figure}[!t]
\centering
\includegraphics[width=0.4\linewidth]{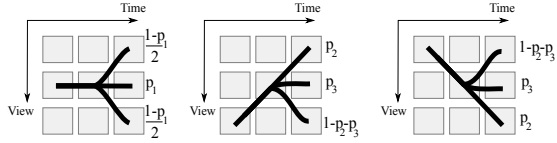}
\caption{Graphical representation of the transition probabilities for  user navigation.}
\label{fig:transProba}
\end{figure}

The navigation model considered in this paper is based on the following intuitive observations.  First, the information of the knowledge of current user position $F_{v,t}$ is not sufficient for predicting the probabilities of choosing the next frames; the system needs to know whether the user is already switching from a view to another one or not. Indeed, let us assume that the user is navigating from left to right, \emph{i.e.,} from $F_{v-1,t-1}$ to $F_{v,t}$, the user will more likely continue switching ($F_{v+1,t+1}$) or remain on the current view ($F_{v,t+1}$) than go back in the other direction ($F_{v-1,t+1}$)\footnote{An identical observation is to be done for a switching from right to left.}. Besides, if the user has been looking at a particular view, he will more certainly continue to display this same view rather than switching to another view (left or right). Based on these observations, we introduce the following transition probabilities:
\begin{align*}
 p(v  | v, v) &= p_1\\
p(v - 1 | v, v) &= p(v + 1 | v, v) = \frac{1 - p_1}{2}  \\
 p(v + 1 | v, v-1) &=   p(v - 1 | v, v+1)  = p_2\\
 p(v  | v, v-1) &=  p(v  | v, v+1) = p_3  \\
 p(v + 1 | v, v+1) &=   p(v - 1 | v, v-1) = 1-p_2-p_3\\
 \end{align*}
where $p(n_1 | n_2,n_3)$ corresponds to the probability that the user chooses the view $n_1$ at time $t$ knowing that he chose the view $n_2$ at $t-1$  and the view $n_3$ at $t-2$. We dropped the time dependency $t$ in the notation for the sake of clarity, as the same transition probabilities are valid at any time $t$. They are graphically represented in Fig.~\ref{fig:transProba}.

These transition probabilities then permit to calculate the popularity of each frame, conditioned on  initial state of the system. Let us assume that at a time $t_0$ a user is displaying the frame $v_0$, and at $t-1$ he was watching the view $v_{-1} \in \{ v_0-1, v_0, v_0+1 \}$. For a request interval $N_T$ and a request delay $N_D$ the set of achievable frames, \emph{i.e.}, the images that can be displayed in the next $N_T$ time instants, is defined by :
$$ \mathcal{F}(F_{v_0,t_0}) = \left \{   F_{v,t_0 + \tau} \ | \  \tau \leq N_T+N_D , \ v_0 - \tau \leq v \leq v_0 + \tau  \right \}. $$
The popularity of each of these frames is calculated as follows :

$\forall~t\geq~t_0, \ \forall v$,
\begin{equation*}
\footnotesize
P(F_{v,t} | F_{v_0,t_0}) = \left\{ 
  \begin{aligned}
  &0 \quad  \text{if} \quad F_{v,t} \notin \mathcal{F}(F_{v_0,t_0}) &\\ 
  &\sum_{\substack{v' = v-1\\F_{v',t-1} \in  \mathcal{F}(F_{v_0,t_0})} }^{v+1} P(F_{v',t-1} | F_{v_0,t_0}) \cdot 
 \Big( \sum_{\substack{v'' = v'-1\\F_{v'',t-2} \in  \mathcal{F}(F_{v_0,t_0})} }^{v'+1} P(F_{v'',t-2} | F_{v_0,t_0}) p(v | v', v'') \Big) \quad \text{otherwise.} &
  \end{aligned}
\right.
\end{equation*}

 In the following, we explain how we use the frame popularities in order to optimize different parts of the general scheme. In particular, we define a rate-distortion efficient coding strategy that gives more importance and typically more bits to the frames that have the highest popularity.

Let us assume that the server has calculated the frame popularity for every image of the future SoF sent at the receiver. The e frame encoding performance can be improved by the allocation of more bits to the frames that have higher chance to be displayed by the user. Based on the probabilities $P(F_{v,t} | F_{v_0,t_0})$ computed earlier, the encoder implements a rate allocation algorithm that adapts the quantization of the residual information in order to minimize the expected distortion at decoder. In other words, the encoder solves a problem of the form 
$$\min_\mathbf{r} \sum_v \sum_t D(\mathbf{r}(v,t)) P(F_{v,t} | F_{v_0,t_0}) \quad \text{s.t.} \quad \sum_v \sum_t \mathbf{r}(v,t) \leq R_{\rm total}  $$ 
where $\mathbf{r}$ is the rate distribution vector limited by a total bit budget $R_{\rm total}$ and $D(\mathbf{r}(v,t))$ is the distortion of the frame at instant $t$ in view $v$, encoded with the rate $\mathbf{r}(v,t)$. As the popularities $P$ do not depend on the rate distribution $\mathbf{r}$, this criterion has a classical form well-known in the rate allocation problem, and thus can be written as:
$$\min_\mathbf{r} \sum_v \sum_t D(\mathbf{r}(v,t)) P(F_{v,t} | F_{v_0,t_0}) + \lambda || \mathbf{r} ||_1 $$ 
where $\lambda > 0$ is the lagrangian multiplier. The resolution of such a problem is simple since it is separable, \emph{i.e.} no dependencies between the distortions $D(\mathbf{r}(v,t))$.

In this allocation problem we focussed on e frames problem. We leave for future works the search of the optimal balance between the rates of depth, reference texture and auxiliary information (as e frames), since it transcendes the scope of the paper. Note that, as the reference frames are used to generate the virtual frames, they are coded with a good quality in order to limit the error propagation in the SoF.


\section{Experimental results}\label{sec:exp}

\subsection{Experimental setup}
The experimental results provided in this section have been obtained with the two sequences of color and depth information provided by Microsoft Research \cite{web_microsoft_ballet_break}, the \emph{ballet} and \emph{breakdancers} sequences (at a resolution of 768 pixels $\times$ 1024 pixels and $15$ frame per second). Both sequences are 100 frames long and contain eight cameras. The rate-distortion (RD) curves correspond to an average of $N_{\rm path}$ experiments  with different user navigation paths. 
The generation of the navigation paths is performed with the same model than the one explained in Sec.~\ref{sec:RA}, which means that the user behavior model used at the server is assumed to match the actual user behavior. We study different aspects of the system performance, like the influence of the system constraints, the storage/bandwidth tradeoff, the role of the user behavior model and the decoding complexity. Most of the experiments have been ran with 10 intermediary views between each of the eight reference views\footnote{This corresponds to the lowest number of view that enables smooth transitions between the reference views.} (otherwise it is specified). The reference views (color and depth) are coded using the scalable mono-view video codec, JSVM \cite{jsvm}. The GOP are synchronized between the views and between the color and depth sequences. The adopted temporal prediction structures in the GOP consider P and B frames. Since all the views are a priori equiprobable, the quantization parameters adopted in the experiments are the same for every views. The e frames are coded independently with the intra mode of JSVM.  We also compare the performance of the proposed systems to baseline solutions.

\subsection{Influence of network constraints}
\begin{figure}[!t]
\centering
\includegraphics[width=0.48\linewidth]{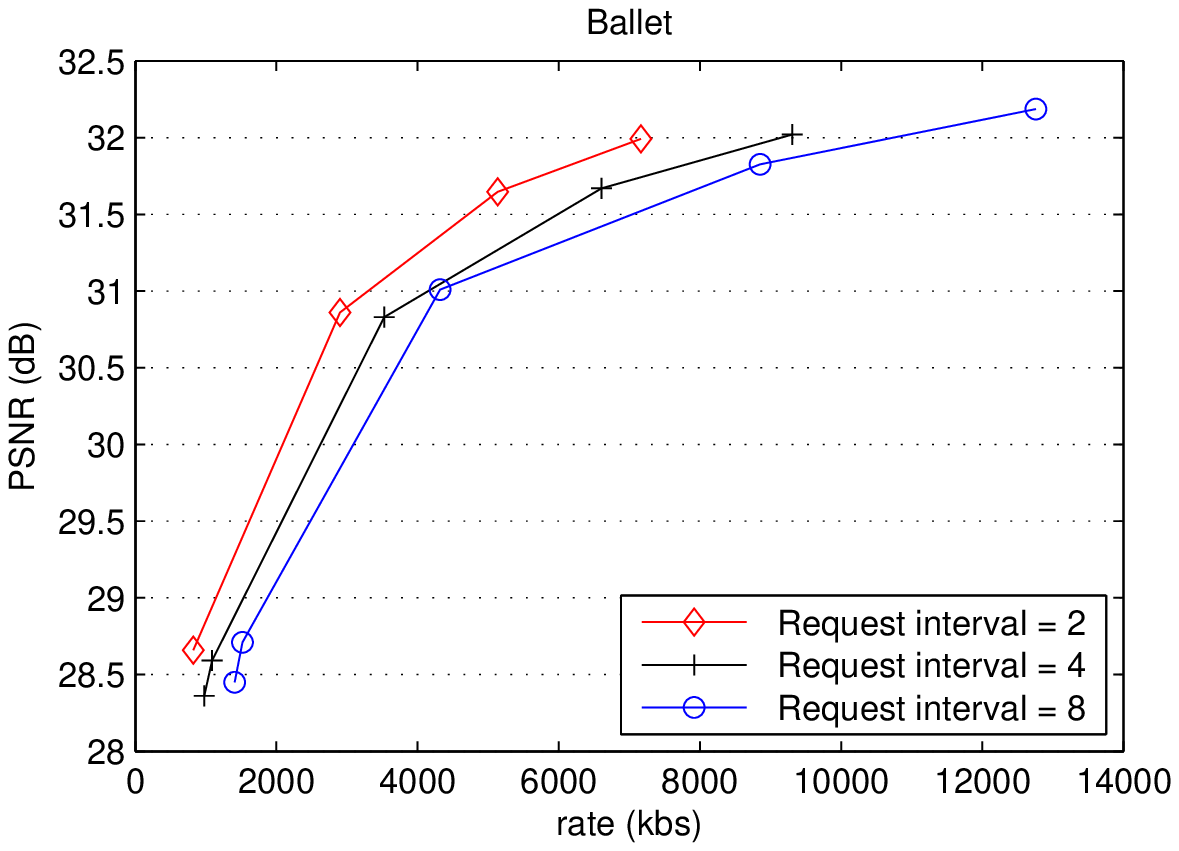}
\includegraphics[width=0.48\linewidth]{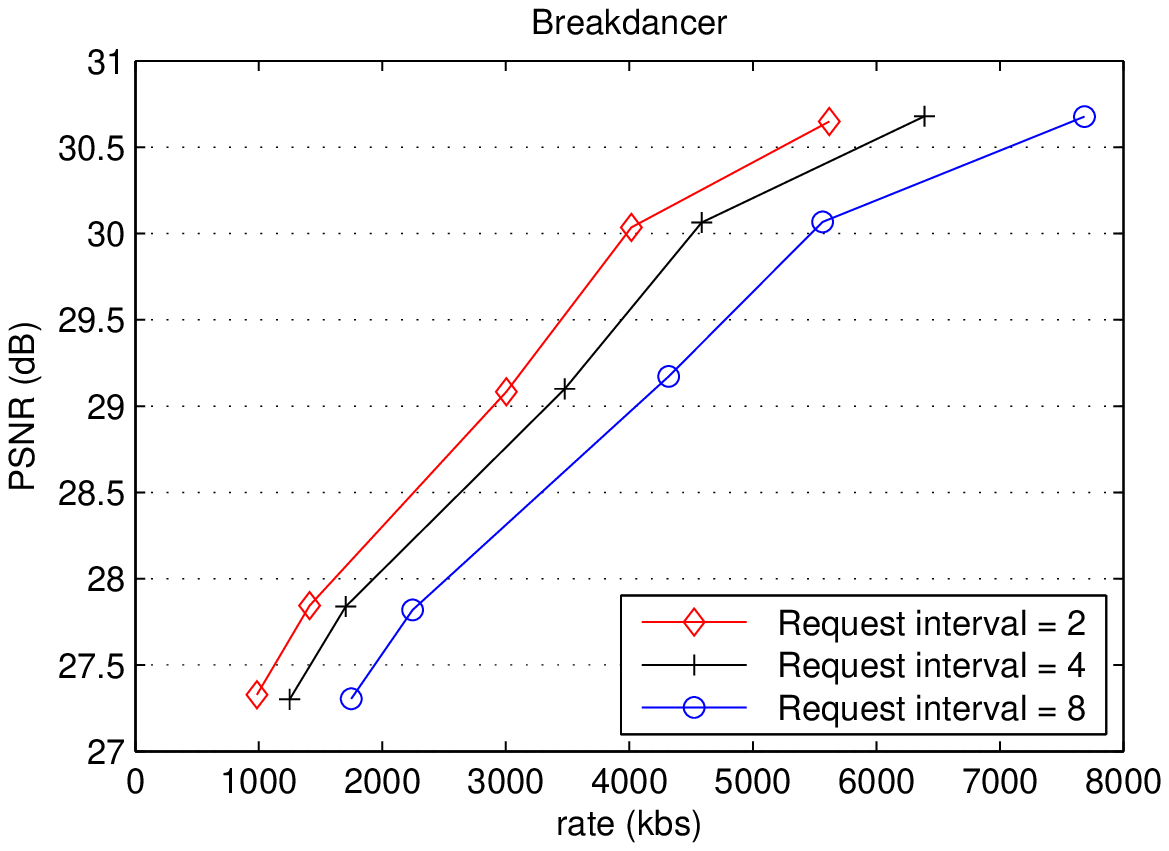}
\caption{RD results for different values of $N_T$  for the \emph{ballet} and \emph{breakdancer} sequences.}
\label{fig:requestInterval}
\end{figure}

As explained in Sec.~\ref{sec:propSystem}, two external system parameters impact on the coding performance. The request interval size $N_T$ corresponds to the level of interactivity allowed by the system. This constraint is often mentioned in the literature but it is not clearly studied.  For example, the authors in \cite{Cheung_G_2011_tip_int_ssmvurfs} consider a scheme with a request interval of 1 and state that, in case of  larger values, the proposed scheme does not permit navigation during the time between two requests. This approach is not conceivable in our case since we want to provide a look around effect to the user. This is why we consider the request delay as an important parameter, and we enable a free navigation between two requests. Therefore the request interval impacts on the number of frames to be transmitted and thus on the quantity of data sent to the user. In Fig.~\ref{fig:requestInterval}, we plot the RD behavior of the system for $N_T$ that is equal to $2$, $4$ and $8$. We observe that the penalty due to large values of $N_T$ is however reasonable. This is explained by the low cost of the e frames that does not significantly impact the system preference when their number  increases. The performance reduction between two configurations can be easily compensated by decreasing the number of intermediate views depending on the target application constraints. Note also that our scheme can adapt to variations of the request interval during the decoding process since no precalculation is performed on the server that precisely depends on this constraint.

\begin{figure}[!t]
\centering
\includegraphics[width=0.48\linewidth]{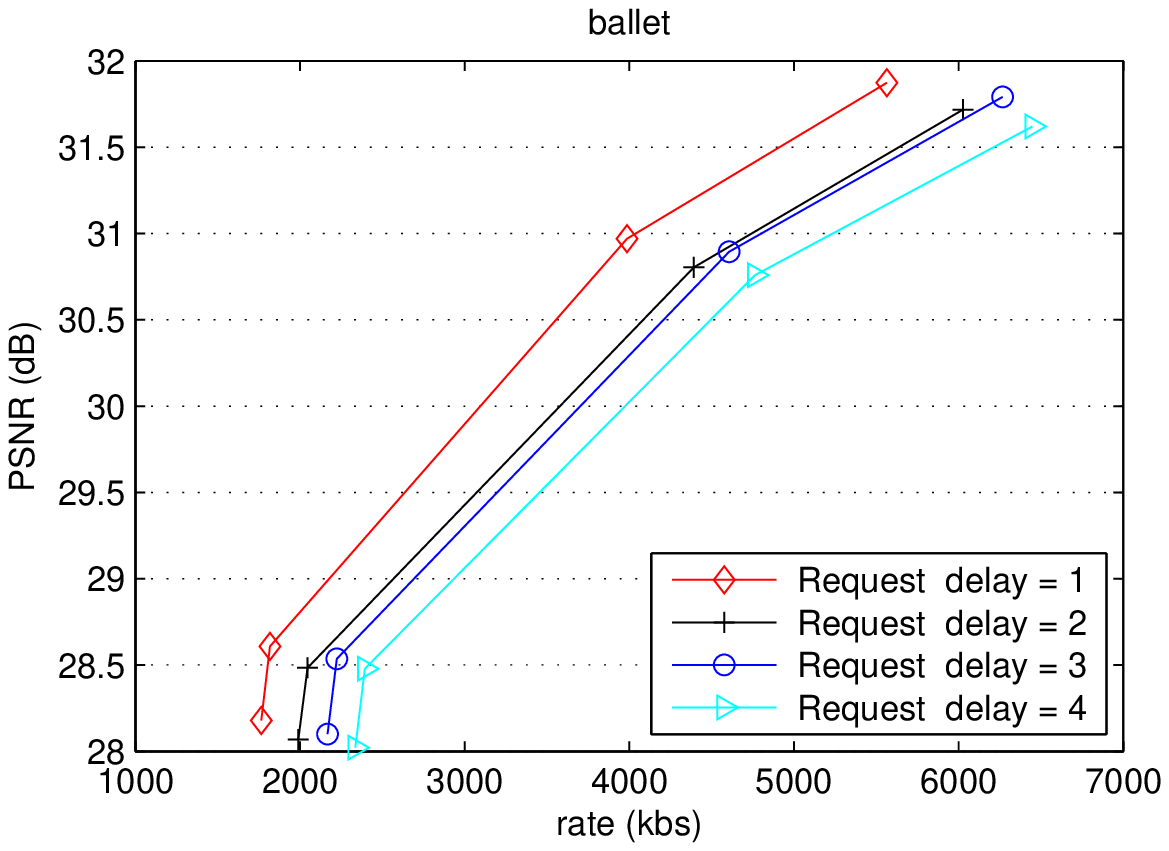}
\includegraphics[width=0.48\linewidth]{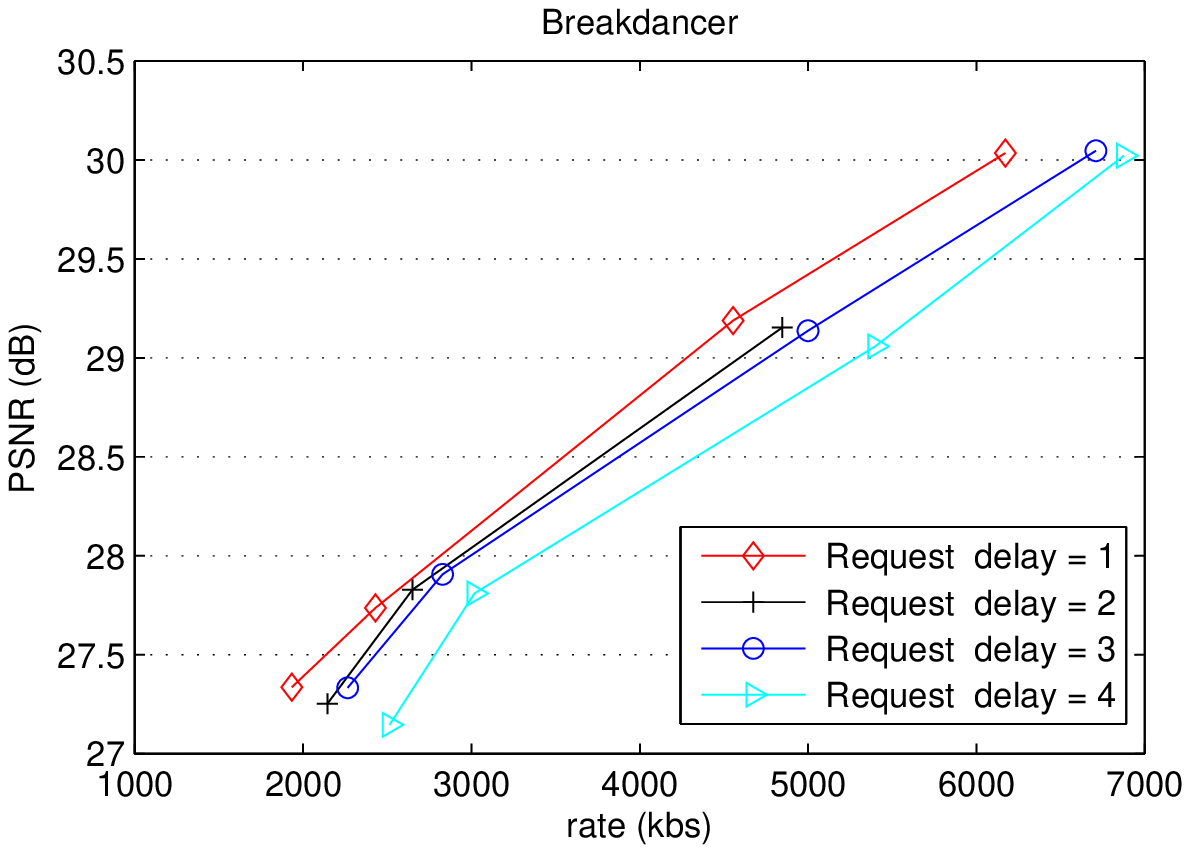}
\caption{RD results for different values of $N_D$  for the \emph{ballet} and  \emph{breakdancer} sequences.}
\label{fig:requestDelay}
\end{figure}

In this work, we also consider the constraint $N_D$ that corresponds to the time between a request and the effective transmission of the corresponding datas. This parameter  depends on the network latency and the time that the server needs to respond to clients' requests. This latter delay could be considerably high for all the methods that  consist in transcoding  or re-encoding the datas in function of the user requests. In our work, everything is prepared and stored on the server beforehand. Then, the response time of the server  is negligible since it only corresponds to the time needed to extract the appropriate bitstream from the scalable description stored on the server. The delay is thus dominated by the network latency. We measure  the influence of the parameter $N_D$  and we show the results in Fig~\ref{fig:requestDelay}. Obviously an increasing value of $N_D$ penalizes the performance, but the consequences are not very important because of the reasonable cost of the e frames. As usual, this performance reduction problem can be handled by design tradeoffs, like decreasing of the navigation smoothness with smaller number of views or by limiting the number of available paths.

\subsection{Compromise between bandwidth and storage}
In a scenario where a user or multiple users simultaneously receive  video sequences, the coding strategy has to deal with a compromise between the storage size on the server and the bandwidth of the transmitted data. A naive scenario in coding all the frames with numerous dependencies and effective prediction with JMVM \cite{jmvm} (\emph{i.e.,} most efficient codec for compressing a whole multiview sequence). However, the coding rate would be tremendous since the display of one frame would require the transmission of numerous other reference frames. This is not efficient in terms of bandwidth. On the opposite, one could consider a situation where the server could stores sequences corresponding to all the possible prediction paths in order to optimize the amount of transmitted data. The storage cost becomes hudge. These two  examples show the intuition that reducing the bandwidth is often obtained by increasing the storage on the server (and \emph{vice versa}) for a given level of interactivity. Some works  (\emph{e.g.,} \cite{Cheung_G_2011_tip_int_ssmvurfs}) aim at finding the coding approach that could give the optimal compromise between storage size and bandwidth. In our work, we do not optimize the prediction structure between the frames. Nevertheless, the tradeoff between bandwidth and storage can be achieved by proper coding of additional information, or by adapting the GOP size of reference views. Since the GOP does not have to be aligned on the value $N_T$, it should be as large as possible for more effective compression, without penalizing delays. However, from a bandwidth point of view, the GOP size should be small in order to reduce the number of  reference frames that are not directly used by the clients. In fact, the optimal GOP size depends on the rate of the intra and predicted images in the reference views. Indeed, if the intra frames are much heavier than the P-frames, the GOP size should be longer in order to reduce the number of I-frames. On the contrary, if the I-frames do not cost too much rate, the GOP should be shorter in order to be adapted to the user navigation. 

\begin{figure}[!t]
\centering
\includegraphics[width=0.5\linewidth]{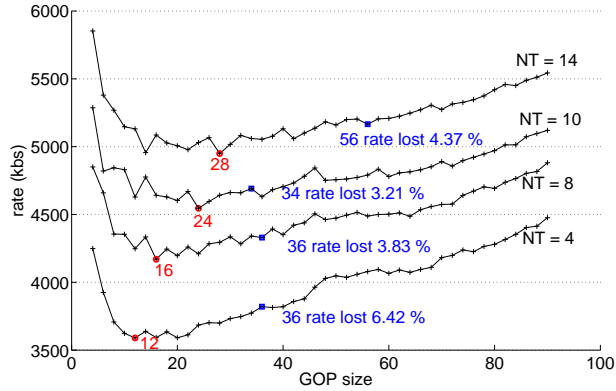}
\caption{Transmission rate for reference views versus GOP size. In red: the optimal GOP size values, when the user behavior model is considered. In blue: the optimal GOP size chosen without user behavior model, along with rate penalty.}
\label{fig:gopSize}
\end{figure}

Finally, another important element to consider in the GOP size selection is the behavior of the user. For example, the GOP size should be short if the user often changes views. Given a user behavior, we find the best GOP size that minimizes the transmission rate without penalizing the compression efficiency of the reference frames. In Fig.~\ref{fig:gopSize}, we show an example of GOP size selection with and without taking the user behavior into account. For a high number of paths, we have simulated the transmission of reference views sequences for different values of $N_T$ and different GOP sizes. For every values of $N_T$ and GOP sizes we have averaged the transmission rate over  different navigation paths, and then we have determined the best GOP size for a given value of $N_T$. In order to analyze the influence of the user behavior model in the decision, we have first determined the GOP size as if all the frames are equiprobable. In a second time, we have generated the path with the transition probabilities defined in Sec.~\ref{sec:RA}. We have compared the bandwidth optimal GOP sizes in both cases. We can observe that the consideration of the user behavior in the selection of the GOP size leads to a rate saving of up to $6\%$ in some situations.

\subsection{RD optimized coding}
\begin{figure}[!t]
\centering
\includegraphics[width=0.38\linewidth]{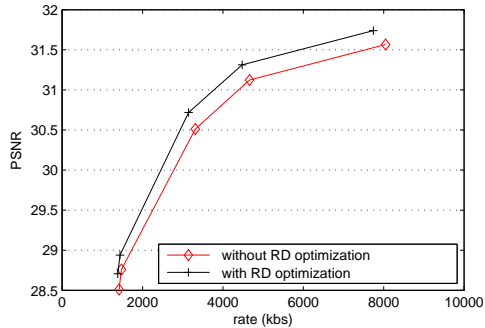}
\caption{Quality versus encoding rate (e frames + reference views)  for \emph{ballet} sequence.}
\label{fig:eAlloc}
\end{figure}

We analyze now the benefits of considering the user behavior model in the rate-distortion optimized  coding of the e frames. We fix the values of $N_T=8$ and $N_D=0$; we transmit the reference and virtual frames such that the view synthesis is performed with two reference images at the receiver, with a block size of $8$. These reference sequences are coded with a GOP size of 16. Fig.~\ref{fig:eAlloc} shows the comparison of the system efficiency with and without the proposed e frame rate allocation introduced in Sec.~\ref{sec:RA}. We can observe that the consideration of the user behavior model in the e frame coding brings a sensible improvement in term of average RD performance compared to an encoding that ignores frame popularities.

We then vary the probabilities $p_1$,  $p_2$ , $p_3$ in the user behavior model of Sec.~\ref{sec:RA};  for each configuration, we measure the performance for the systems with and without  optimized rate allocation, while the decoding process follows the user behavior model. We propose four realistic situations that could be seen on actual user navigation processes. The results are presented in Tab.~\ref{tab:inflProbModel}. The first remark concerns the very interesting gain that we obtain for every scenario when encoding is optimized by considering a user behavior model. In all the cases, taking into account the user behavior model leads to a non-negligible rate saving greater than 9\% (in terms of Bjontegaard metric \cite{Bjontegaard_G_2001_tr_cal_apsnrdbrdc}). Moreover,  the variation of the gain between scenarios is interesting to analyze in detail. In the situation where the user performs a random navigation, the gain is less important than in the situation where the navigation is almost deterministic (first line). In real situations, it is obvious that the user behavior would follow different modes depending on the scene content. Our results show that  our rate allocation solution  leads to interesting rate-distortion performance even with a more evolved probabilistic model that could detect the different navigation modes.

\begin{table}
\centering
\begin{scriptsize}
\begin{tabular}{c|c|c|c|c|c}
 $p_1$  &  $p_2$  & $p_3$   & type of trajectory & in practice & Rate saving (Bjontegaard \cite{Bjontegaard_G_2001_tr_cal_apsnrdbrdc})  \\ \hline
 0.9 &  0.1  & 0.9   &   almost no switching & the user remains on a nice viewpoint  & -13 \% \\
 0.3  &  0.3  & 0.3   &  almost random navigation &  the user is looking for the best viewpoint  & -9 \%\\
   0.1  & 0.9   & 0.1   & long switch in the same direction & the user completely changes the viewpoint  & -10 \% \\
    0.1 & 0.1   & 0.1   & zigzag & the user tests the look around effect  & -10 \% \\
 \hline   \end{tabular}
 \end{scriptsize}
  \caption{Influence of the probability model on the Bjontegaard gain between the cases with and without model-based rate allocation.}
  \label{tab:inflProbModel}
   \end{table}

\subsection{Decoding complexity}

We now analyze the performance of our system in terms of computational complexity at decoder, which was one of the main motivations for the construction of e frames. The machine used for these experiments is a quad cores, Intel(R) Xeon(R) (2.66 GHz). We consider in these experiments that the network delay, $N_D$, is zero. In the first column of Tab.~\ref{tab:complexReduct}, we present the computational time savings of the proposed low-complexity VVS algorithm (block-based disparity compensation and  summation of  residual information). We consider different block size configurations (1, 4, 8 and 16 pixels) in the disparity compensation and we calculate the computational time savings in our decoder with respect to the complex VVS techniques involved in the classical decoding schemes. The  results demonstrate that our scheme leads to computational complexity savings that are really significant. The second column shows the complexity reduction for the whole decoding process, \emph{i.e.,} the reference and e frames decoding processes. The complexity reduction results are pretty convincing about the interest of transmitting additional information as the e frames in interactive multiview systems. 

\begin{figure}[!t]
\centering
\begin{minipage}{0.48\linewidth}
\includegraphics[width=0.95\linewidth]{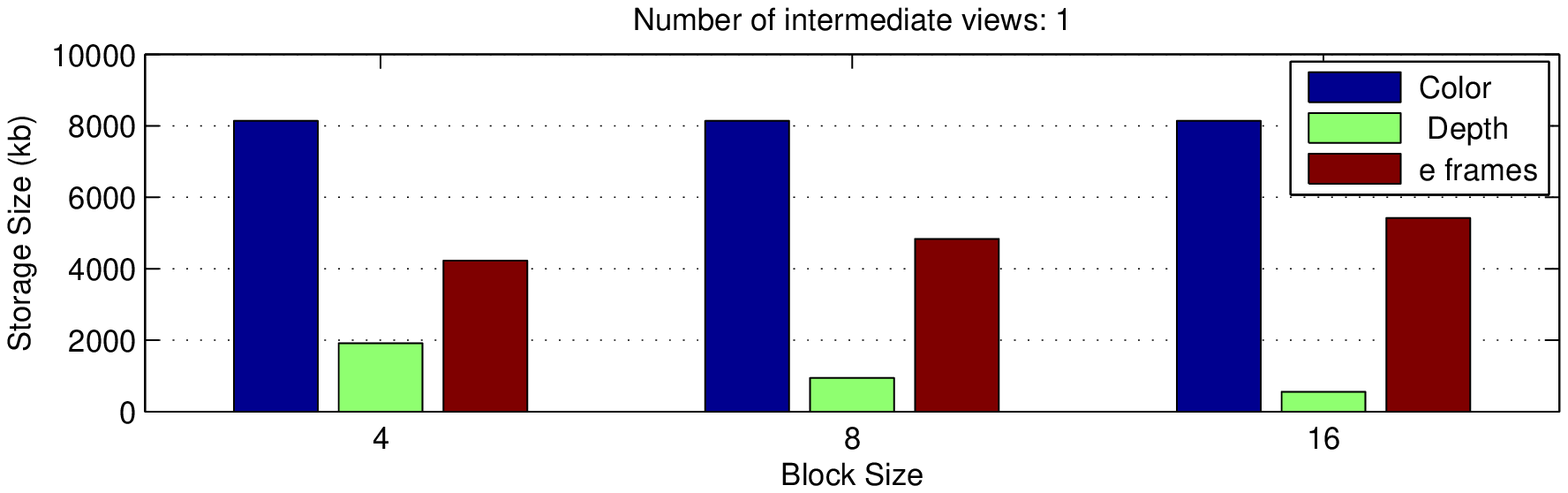}
\includegraphics[width=0.95\linewidth]{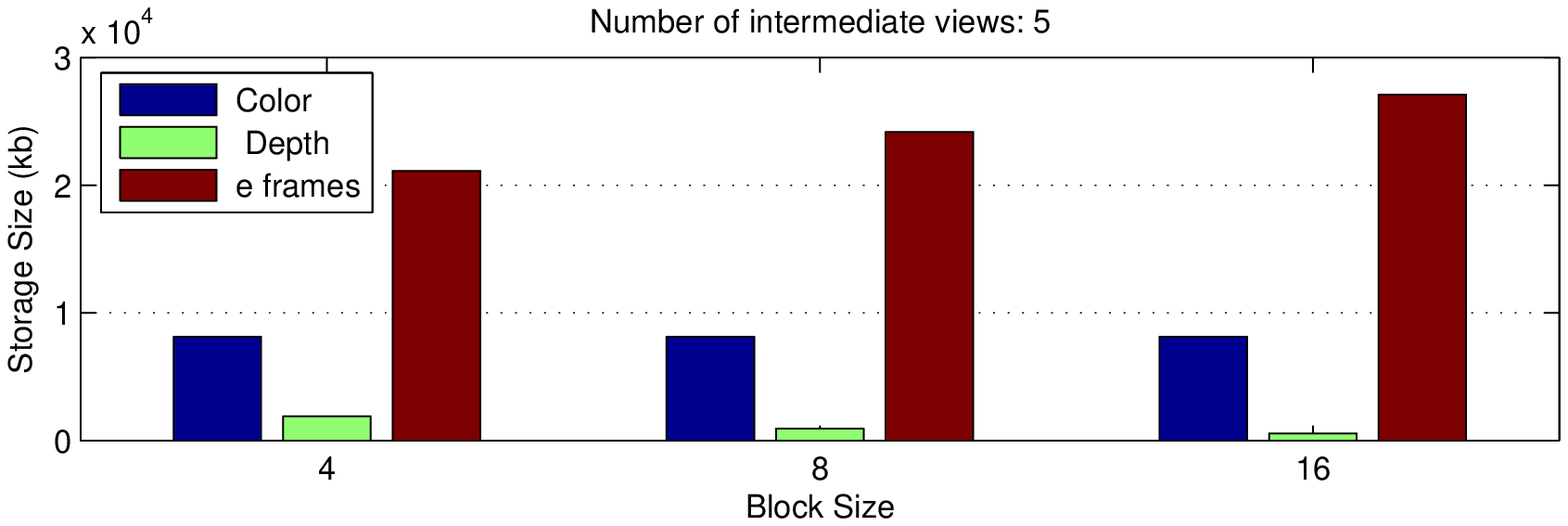}
\includegraphics[width=0.95\linewidth]{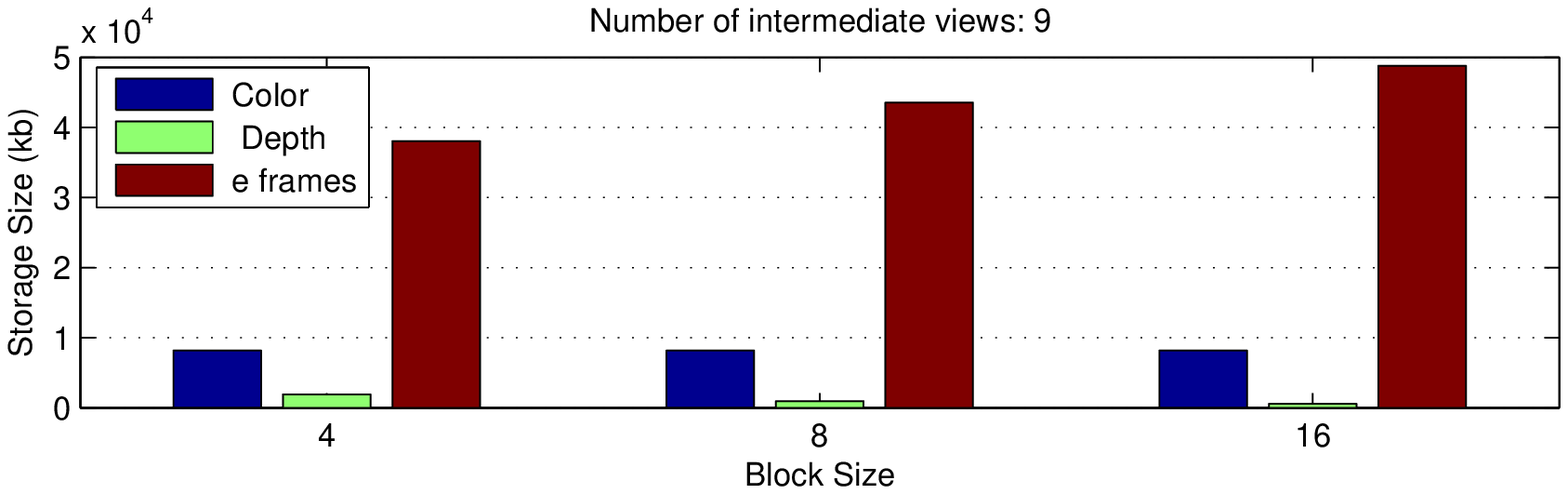}
\caption{Storage size on the server as a function of the disparity compensation block size $B$.}
\label{fig:StorageSize_blckSize}
\end{minipage}
\begin{minipage}{0.48\linewidth}
\includegraphics[width=1\linewidth]{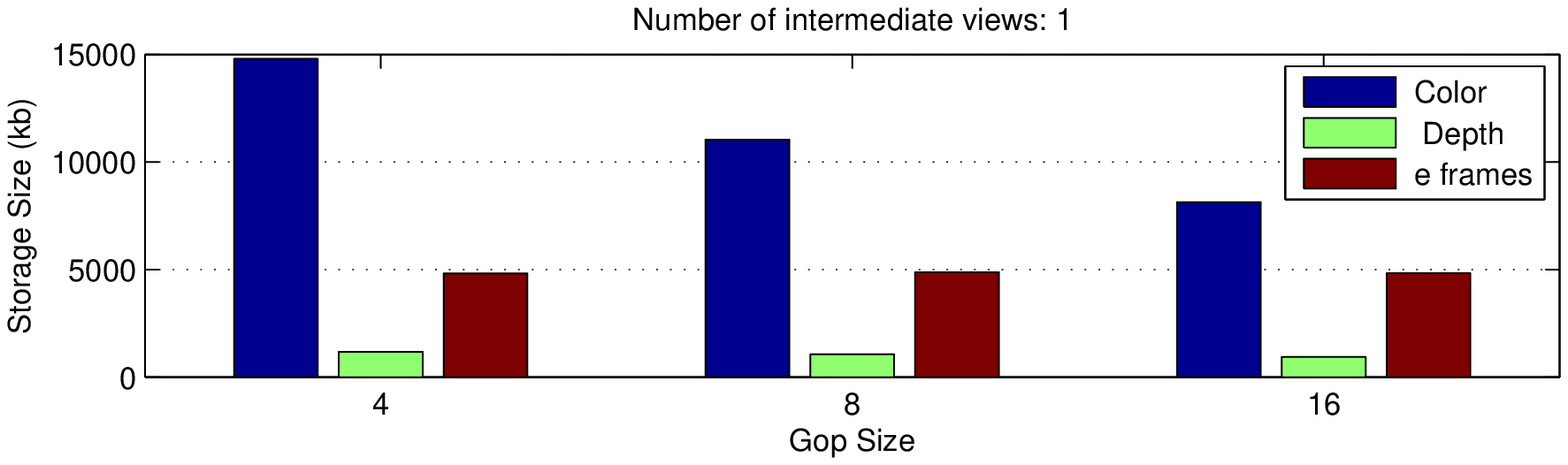}
\includegraphics[width=1\linewidth]{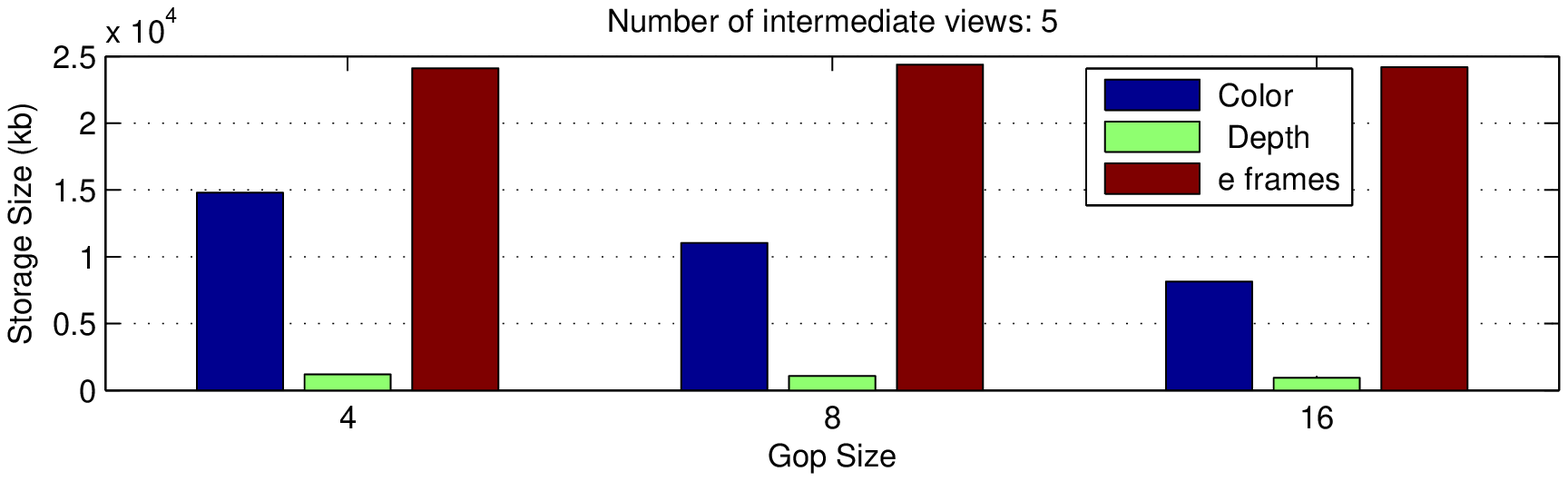}
\includegraphics[width=1\linewidth]{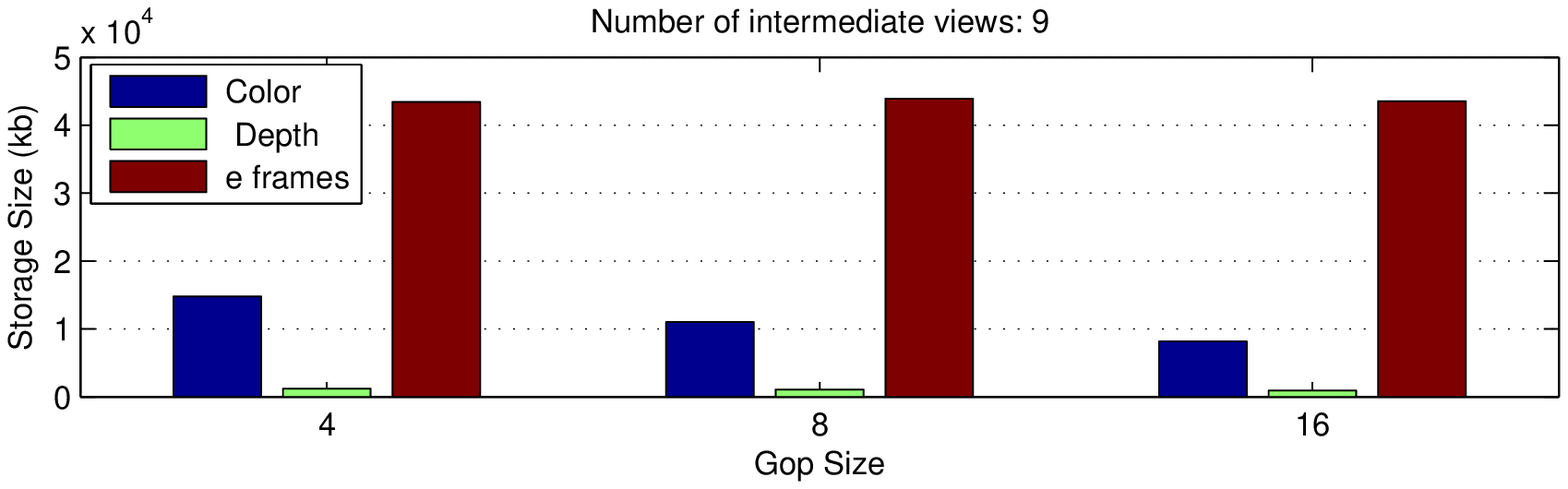}
\caption{Storage size on the server as a function of the GOP size in coding the reference views.}
\label{fig:StorageSize_gopSize}
\end{minipage}

\end{figure}

This considerable decoding time reduction does however not come for free as the third column of Tab.~\ref{tab:complexReduct} shows it, since the variance of the residual information increases with $B$. The effective cost of the e frames is shown in Figs.~\ref{fig:StorageSize_blckSize} and \ref{fig:StorageSize_gopSize}. These figures represent the storage sizes on the server of the three following entities: the reference color sequence, the depth images, and the e frames. In Fig.~\ref{fig:StorageSize_blckSize} (resp Fig.~\ref{fig:StorageSize_gopSize}), the evolution of these quantities is given in function of $B$,  VVS block size $B$ (resp. the GOP size of the reference frames) and in function of the number of intermediate views  that we considered between the reference views.  One can see that the e frames storage cost is not negligible but remains reasonable considering the number of virtual views that they can generate. For instance, the e frame size is slightly higher than twice the size of the color image reference, whereas they permit to generate $10$ times more views, which considerably improves the smoothness of the navigation to produce the look around effect. It is also important to note that the storage size does not exactly correspond to the transmission rate. The cost of the e frames during the transmission  between the server and a client is given in Tab.~\ref{tab:rateRepartition}. In these experiments, we transmit all the information needed for an interactive navigation at the receiver, and we measure, at medium bitrate, the weight of each entity (reference color, reference depth and e frames) in the total bit budget. We still assume here that the network delay, $N_D$, is zero.  Although this cost is not negligible, it remains inferior to one third of the total bit budget in the case of very smooth view transitions, (\emph{i.e.,} 10 intermediary views). If this cost is too high for given bandwidth constraints, one can reduce the smoothness of the navigation and consider a smaller number of intermediary views. For example, the results in Tab.~\ref{tab:rateRepartition2} show that, when the number of intermediary views is set to $5$, the relative rate of e frames is sensibly reduced and never increases beyond $\frac{1}{5}$ of the total bitrate. 

Overall, the experiments shown in this section demonstrate that our scheme manages to provide a considerable complexity reduction with respect to the existing decoding schemes for interactive multiview navigation. The cost of this low complexity decoding is reasonable and further reducible by adapting the interactivity and navigation quality levels.

\begin{table}
\centering
\begin{scriptsize}
\begin{tabular}{c | c | c | c }
 Configuration  &  \begin{minipage}{0.2\linewidth}\center \tiny Computational time reducing of our VVS technique (projection + summation of a residual)  wrt to the complex VVS algorithm (dense projection + inpainting) \end{minipage}&  \begin{minipage}{0.2\linewidth}\center \tiny Frame decoding time reducing of our system (projection + residual decoding + summation of the residual)  wrt to the complex decoding approach (dense projection + inpainting) \end{minipage}    & \begin{minipage}{0.2\linewidth}\center \tiny Variance of the residual \end{minipage} \\ \hline
dense projection  & 2.20 \% & 4.24\% & 224.48 \\
$B = 4\times4$  & 0.16\% & 1.42\% & 273.71 \\
$B = 8\times8$  & 0.00047\% & 0.81\% & 292.88\\
$B = 16\times16$  & 0.00017\% & 0.61\% & 333.02 \\
 \hline   \end{tabular}
 \end{scriptsize}
  \caption{Calculation time reducing for the VVS algorithm (second column) and the whole decoding process (third column) for different values of the block size $B$. The third column corresponds to the variance of the residual information.}
  \label{tab:complexReduct}
   \end{table}

\begin{table}
\centering
\begin{scriptsize}
\begin{tabular}{c | c | c | c }
 Configuration  &  color rate (\%)  & depth rate (\%)  & e frames rate (\%)  \\ \hline
$B = 4\times4$ + $GOP$ 16 & 61.2 & 18.7 & 20.1  \\
$B = 8\times8$ + $GOP$ 16 &  68.0 & 10.5 & 21.5 \\
$B = 16\times16$ + $GOP$ 16 & 68.5 & 6.1 & 25.4\\ \hline
$B = 4\times4$ + $GOP$ 8  &56.7 & 17.4 & 25.9 \\%
$B = 8\times8$ + $GOP$ 8  & 64.9 & 9.0 & 26.0 \\
$B = 16\times16$ + $GOP$ 8  & 62.4 & 6.4 & 31.2  \\%
 \hline   \end{tabular}
 \end{scriptsize}
  \caption{Rate distribution in a system with 10 intermediary views ($N_T=2$).}
  \label{tab:rateRepartition}
   \end{table}

\begin{table}
\centering
\begin{scriptsize}
\begin{tabular}{c | c | c | c }
 Configuration  &  color rate (\%)  & depth rate (\%)  & e frames rate (\%)  \\ \hline
$B = 4\times4$ + $GOP$ 8  &  65.4 & 20.5 & 14.1 \\%
$B = 8\times8$ + $GOP$ 8  & 72.4 & 11.3 & 16.4 \\
$B = 16\times16$ + $GOP$ 8  &  73.1 & 7.4 & 19.5  \\%
 \hline   \end{tabular}
 \end{scriptsize}
  \caption{Rate distribution in a system with 5 intermediary views ($N_T=2$).}
  \label{tab:rateRepartition2}
   \end{table}

\subsection{Comparisons with other solutions}
\begin{figure}[!t]
\centering
\begin{minipage}{0.78\linewidth}
\includegraphics[width=0.98\linewidth]{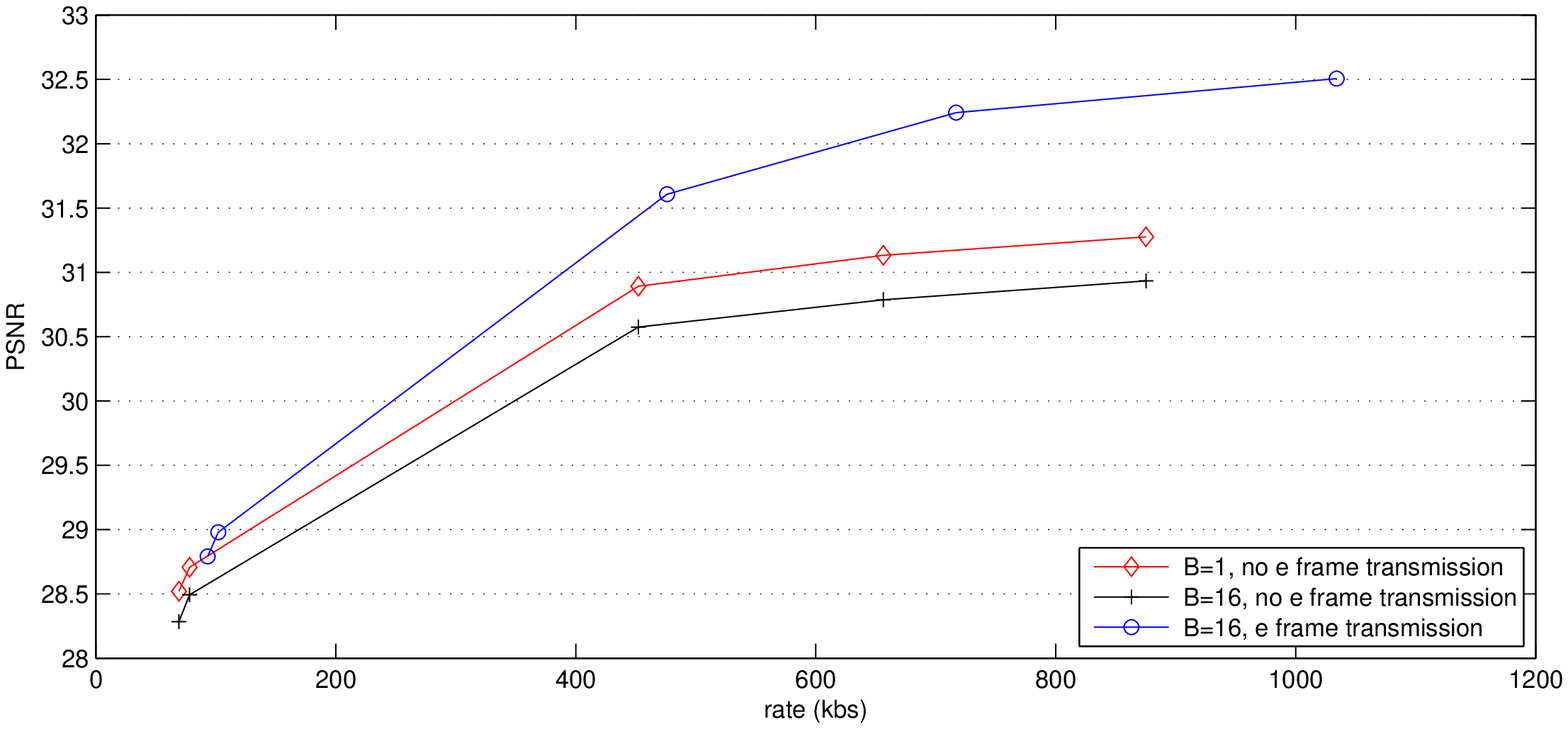}
\centerline{\footnotesize (a) \emph{ballet}}
\end{minipage}
\begin{minipage}{0.78\linewidth}
\includegraphics[width=0.98\linewidth]{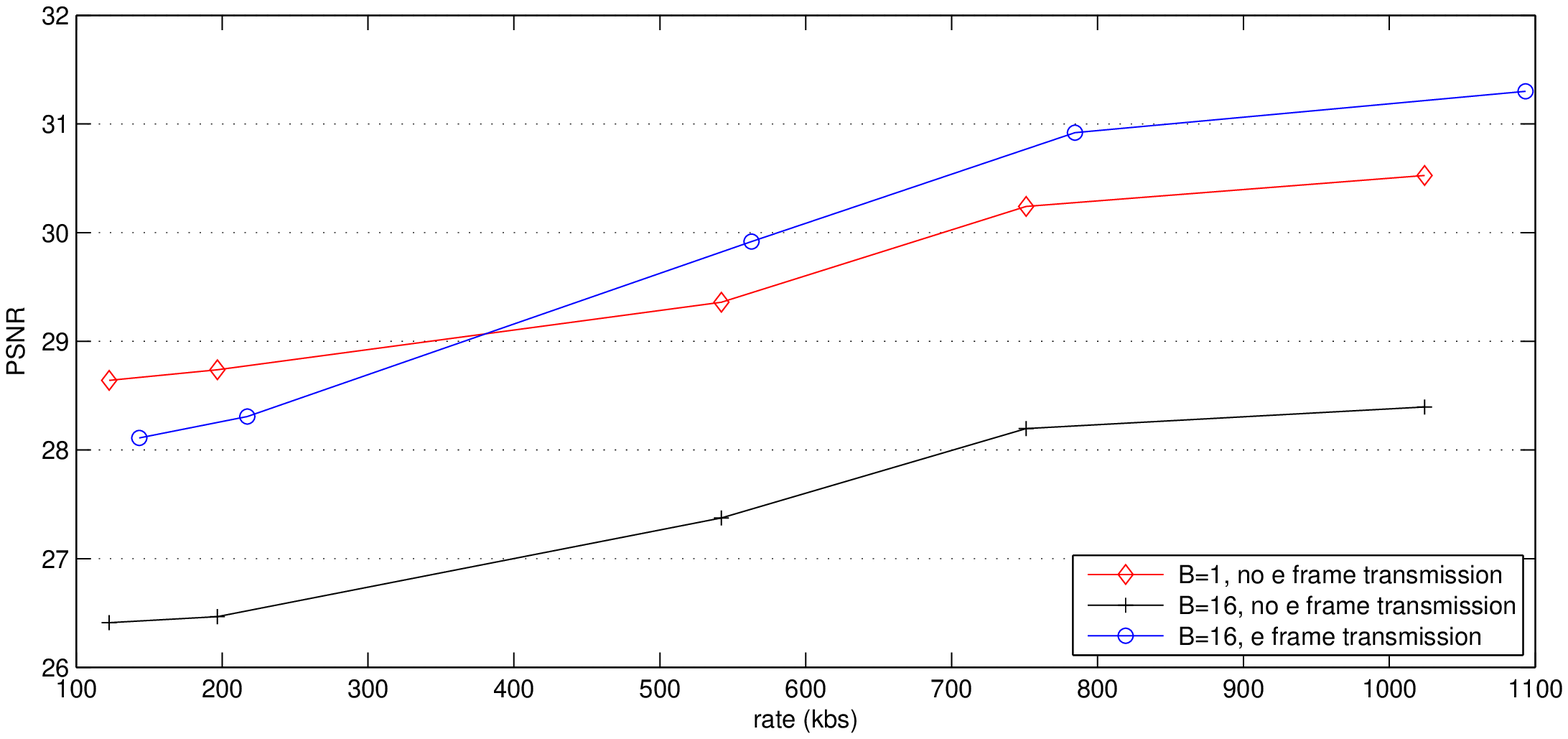}
\centerline{\footnotesize (b) \emph{breakdancer}}
\end{minipage}
\caption{Comparison of decoding performance when e frames are transmitted or not.}
\label{fig:noEcoding}
\end{figure}

Our proposed system is  a complement to the existing schemes rather than an completely different alternative, since it explores the virtual view synthesis with a low-power decoder assumption.  Nevertheless, we present in Fig.~\ref{fig:noEcoding} some tests that provides hints about the benefits of the e frames solution. In these experiments, we use the following coding parameters: $N_T=8$, $N_D=0$, GOP size of 8. The e frames are transmitted using the RD optimized coding strategy. The proposed scheme (blue curve, squares) corresponds to a configuration with the block size $B=16$. In order to measure the importance of e frames in the reconstruction quality, we plot the curves corresponding to the situation where the block size is identical (black line and crosses), but where the e frames are not transmitted and rather replaced at the decoder by a simple inpainting method (averaging of the neighboring pixels). This alternative system mimics the behavior of a decoder that cannot afford medium to high complexity in VVS. The resulting curves clearly highlight the benefits of  e frames in terms of visual quality. Another interesting comparison is to consider that the decoding process is a bit more powerful and is able to calculate a dense projection ($B=1$) with a similar inpainting than in the previous scheme (without e frame transmission). The results (represented in red line, losange) shows that for medium and high bitrate, it is worth sending residual information rather than having a very precise projection and bitrate savings. For lower bitrate, the relative performance is sometimes different. This is explained by the fact that the cost of the e frames is proportionally higher at low bitrate, exactly like the motion vectors in classical video coding.

\section{Related work}\label{sec:soa}
Our work  serves as a complement to the current literature that tackles the interesting problem of providing interactivity in multi-view video coding or streaming. We see in this section that the existing methods address the problem of reference view transmission while our system rather studies the question of sending information to help the virtual view synthesis (not considered in the techniques detailed below). We review in this section the most relevant works that address the design of interactive video services.

The introduction of interactivity in video systems has first been explored for mono-view video where the problem consists in enabling the user to access every frame in the sequence with a minimum delay. With the classical coding schemes (\emph{e.g.}, H.264) if a user accesses a frame randomly, and if this frame is a predicted frame, the decoder has to receive and decode a set of intermediary frames, which leads to a non-negligible decoding delay. It requires the transmission of useless frames with a penalty in rate-distortion performance. Some solutions have been proposed in order to tackle this problem. One of them is  based on  SI and SP frames \cite{Karczewicz_M_2003_tcsvt_sp_sifdh}, which are images added in H.264 bitstream that help for switching between two bitstreams or for random access. These SP/SI frames are constructed with motion prediction with reasonable encoding sizes. This solution is then less costly than simple solutions that transmit intra frames at the switching instants. Another technique \cite{Cheung_NM_2006_pvicip_vid_cfpobdsc} uses a similar idea of building  predicted frames that do not depend on the reference image they are predicted from. It uses distributed source coding techniques and transmits hash information in order to construct the side information at the Slepian-Wolf decoder.
The solutions proposed to solve the problem of providing interactivity in mono-view scenario lays the foundations of a general problem of adapting the encoding strategy to the user behavior. The general idea is to anticipate the user behavior with two possible alternative: i) to send additional information or ii) construct a complete prediction structure between the images.

\subsubsection{Multi-view view switching}

A straightforward extension of mono-view interactivity to multiview systems has been proposed in \cite{Chen_Y_2009_jadvsp_eme_mvcstdvs}, which adapts the concept of SP/SI frames to view switching. As in mono-view system, these frames constitutes additional information to help the transition between two predefined GOPs. While this approach is appropriate in the case of mono-view video (since the user does not switch too often), it becomes limited for view switching because the user may change  the displayed viewpoint frequently, which requires a high quantity of  additional information with SP/SI frames. Another approach has been proposed in \cite{kurutepe_E_2007_tcsvt_cli_dssmvitdtv} and reviewed in \cite{Tekalp_AM_2007_ieee-spm_tdt_oipesmv}. It consists in describing the signal in different layers with different levels of prediction. In other words, the encoder provides different descriptions of the signal that can enhance the frame reconstruction when the user changes the viewpoint. The user position is predicted using a Kalman filter. The authors in \cite{Liu_Y_2010_jvci_rdo_ismvme} alternatively propose to store multiple encodings on the server and to adapt the transmission to the user position. This however brings a high storage cost on the server.

As in the mono-view scenario, some other works adapt the prediction structure to the user behavior. In  \cite{Lou_JG_2005_real_timvvs} the system performs real-time encoding and enables the user to switch at precise instants (when the target frame is intra coded). To tackle the limitation of real-time encoding, other works have been proposed such as in \cite{Kimata_H_2004_ntt_fre_vvcumvc,Shimizu_S_2007_tcsvt_vie_smvcutdwdm} where the multi-view sequence is encoded with a GoGOP structure, that corresponds to a set of GOP. Inside a GoGOP the frames are coded using different predictions in order to preserve the compression efficiency. On the other hand, the GoGOP are coded independently in order to enable view switching without transmitting large sets of useless frames. The limitation of such methods in the fixed encoding structure, which cannot be easily adapted to different configurations. In some situation, the user may indeed change viewpoints more frequently than in other cases. Interested readers may refer to \cite{Smolic_A_2005_pieee_int_tdvrct} and  \cite{Tekalp_AM_2007_ieee-spm_tdt_oipesmv}  that give a good overview of these interactive multiview decoding techniques. The first work that provides an optimization of the prediction structure for interactive decoding has been developed in \cite{Chaung_2008_mmsp_cod_soimsvwo,Cheung_G_2009_ipvw_gen_rfsims}. The problem is formulated so that the proposed prediction structure reaches a compromise between storage and bandwidth. The possible type of frames are intra frames and predicted frames (with the storage of different motion vectors and residuals). Petrazzuoli \emph{et al.} \cite{Petrazzuoli_G_2011_picip_usi_dscdibriimva} have recently introduced the idea of using distributed source coding and inter-view prediction for effective multiview switching.

Both ideas of adapting the frame prediction structure and creating additional information have been merged in \cite{Cheung_G_2009_picip_opt_fsudscimvs,Cheung_G_2011_tip_int_ssmvurfs} that extend the work in \cite{Chaung_2008_mmsp_cod_soimsvwo,Cheung_G_2009_ipvw_gen_rfsims} by adding another possible frame description type based on distributed source coding techniques. This has been recently extended in \cite{Xiu_X_2011_picip_fra_soimvsbnd} by taking into account the network delay constraints. With this approach the description of the multiview sequence becomes quite efficient, but this solution does not deal with the question of view synthesis. The scheme  proposed in this paper offers a complementary solution to such techniques.

Finally, another important issue in multiview video streaming is the design of systems that enable the transmission of 3D information to multiple heterogeneous users with data representation described above. The purpose of these systems is to meet users' requests under different constraints (\emph{e.g.,} delay, bandwidth, power resources, etc.). In this context, only a few works address at the decoder the problem posed by the limitation of computational complexity during the view rendering at the decoder. In \cite{Shi_S_2009_pacmicm_rea_trrtdvmd}, the system contains intermediary servers that performs the virtual view rendering in the place of the light decoders and transmit the resulting images to decoders. However, this approach leads to high processing delays which can be addressed by choosing the appropriate remote rendering systems \cite{Shi_S_2010_pacmicm_hig_qldrrs3dv}. The SyncCast system \cite{Huang_Z_2011_pacmicm_syn_sdmsitdti} moreover enables  the user to interact with each other for improved decoding performance. All of these works can lead to interesting extensions of our solution, where only one server currently delivers the video sequence to multiple users. The format of the data that we have considered moreover considerably reduces this processing delay because everything can be distributed to multiple servers.

\section{Conclusion}

In this paper we have studied the question of reducing the required power (or increasing battery lifetime) at the receiver side of an interactive multi-view video coding system. Our original idea consists in sending residual frame information that helps smooth view navigation at the decoder. We have shown that the cost of this additional information is reasonable and that it can be even reduced by integrating the user behavior in effective rate allocation strategies. Our work interestingly provides a system that could be readily  implemented on the nowadays decoding devices. Finally, it introduces the idea of sending residual information for virtual views, which could trigger some future research work with additional purposes such as the implement of the compression efficiency.

\ifCLASSOPTIONcaptionsoff
  \newpage
\fi

\bibliographystyle{IEEEtran}
\bibliography{abbr,epfl}

\end{document}